\documentclass[aps,groupedaddress,prb,twocolumn,showpacs,superscriptaddress]{revtex4}
\usepackage{graphicx}
\usepackage{amsmath,amssymb,amsfonts}
\usepackage{natbib}

\newcommand{\be}{\begin{equation}}
\newcommand{\ee}{\end{equation}}
\newcommand{\ba}{\begin{eqnarray}}
\newcommand{\ea}{\end{eqnarray}}

\begin{document}

\title{Two-bath model for activated surface diffusion of interacting
adsorbates}

\author{R. Mart\'{\i}nez-Casado}
\email{r.martinezcasado@imperial.ac.uk}
\affiliation{Department of Chemistry, Imperial College,
London SW7 2AZ, United Kingdom}

\author{A.S. Sanz}
\email{asanz@imaff.cfmac.csic.es}
\affiliation{Instituto de F\'{\i}sica Fundamental,
Consejo Superior de Investigaciones Cient\'{\i}ficas,
Serrano 123, 28006 Madrid, Spain}

\author{G. Rojas-Lorenzo}
\email{grojas@imaff.cfmac.csic.es}
\affiliation{Instituto de F\'{\i}sica Fundamental,
Consejo Superior de Investigaciones Cient\'{\i}ficas,
Serrano 123, 28006 Madrid, Spain}
\affiliation{Instituto Superior de Tecnolog\'{\i}as y Ciencias
Aplicadas, Ave.\ Salvador Allende y Luaces, Quinta de Los Molinos,
Plaza, La Habana 10600, Cuba}

\author{S. Miret-Art\'es}
\email{s.miret@imaff.cfmac.csic.es}
\affiliation{Instituto de F\'{\i}sica Fundamental,
Consejo Superior de Investigaciones Cient\'{\i}ficas,
Serrano 123, 28006 Madrid, Spain}

\date{\today}

\begin{abstract}
The diffusion and low vibrational motions of adsorbates on surfaces
can be well described by a purely stochastic model, the so-called
{\it interacting single adsorbate model}, for low-moderate coverages
($\theta\lesssim 0.12$).
Within this model, the effects of thermal surface phonons and
adsorbate-adsorbate collisions are accounted for by two uncorrelated
noise functions which arise in a natural way from a two-bath model
based on a generalization of the one-bath Caldeira-Leggett Hamiltonian.
As an illustration, the model is applied to the diffusion of Na atoms
on a Cu(001) surface with different coverages.
\end{abstract}

\pacs{05.10.Gg, 05.40.-a, 68.43.-h}

%02.50.Ey Stochastic processes
%05.10.Gg Stochastic analysis methods (Fokker-Planck, Langevin, etc.)
%05.40.-a Fluctuation phenomena, random processes, noise, and Brownian motion
%68.43.-h Chemisorption/physisorption: adsorbates on surfaces
%82.20.Db Transition state theory and statistical theories of rate constants

\maketitle

%%%%%%%%%%%%%%%%%%%%%%%%%%%%%%%%%%%%%%%%%%%%%%%%%%%%%%%%%%%%%%%%%%%%%%%
%%%%%%%%%%%%%%%%%%%%%%%%%%%%%%%%%%%%%%%%%%%%%%%%%%%%%%%%%%%%%%%%%%%%%%%

\section{Introduction}
\label{sec1}

Nowadays, quasielastic Helium atom scattering\cite{hofmann,salva}
(QHAS),  $^3$He spin-echo\cite{allison} and neutron
spin-echo,\cite{farago,fouquet} are well established experimental
techniques to study fast diffusion processes which are aimed at
determining gas-surface interaction potentials.
The observables that can be measured with these techniques are the
so-called {\it dynamic structure factor} and the {\it intermediate
scattering function} (or {\it polarization function}), which is the
inverse time Fourier transform of the former.
The diffuse elastic intensity of the He atoms scattered away at large
angles from the specular direction provides very detailed information
about the mobility of adsorbates on surfaces.
Based on the transition matrix formalism, and applied to the first
technique, Manson and Celli\cite{manson} proposed a quantum diffuse
inelastic theory for small coverages of adsorbates on the surface,
ignoring multiple scattering effects with the incoming He atoms.
From this theory, they obtained the dynamical structure factor under
the assumption that all crystal vibrational modes and point-like
scattering centers satisfy the harmonic approximation with a given
frequency distribution.

Alternatively, the study and analysis of surface diffusion processes
can also be tackled by means of the Langevin formalism in two
dimensions.
This assumption is reasonable because, usually, the adparticle motion
normal to the surface involves vibrational modes with much higher
frequencies than in-plane vibrations.
Nonetheless, some recent work shows\cite{allison} that motion
perpendicular to the surface, related to a translational hopping
diffusion process, could be important for coverages $\theta \gtrsim
0.05$, as noticed for small parallel momentum transfers.
As is well known, one of the most popular and ubiquitous
phenomenological equations for systems interacting with environments
is the generalized Langevin equation (GLE).
In the case of surface diffusion, this equation reads as
\begin{equation}
 \ddot{{\bf R}}(t) = - \int_0^t {\bf K} (t-t') \ \!  \dot{{\bf R}}(t')
  \ \! dt' + {\bf F}[{\bf R}(t)] + \delta {\bf F}_f (t) ,
 \label{GLE}
\end{equation}
where ${\bf R} = (x,y)$ is the adsorbate position at a certain
equilibrium distance upon the surface and
${\bf F} = - \nabla V$ is the deterministic force acting on the
adsorbate, with $V$ being the periodic adsorbate-surface interaction
potential determined from He scattering in the zero coverage and zero
surface temperature limits.\cite{graham1}
Equation~(\ref{GLE}) is a stochastic integro-differential equation
with additive fluctuations and linear dissipation, for neither the
fluctuating force, $\delta {\bf F}_f = {\bf F}_f - \langle {\bf F}_f
\rangle$, nor the dissipative kernel, ${\bf K}(t-t')$, depend on the
adsorbate position ${\bf R}$.
Usually, the fluctuating force is considered as a zero-centered
Gaussian, completely specified by its autocorrelation function,
${\bf C}(t-\tau)$, which is assumed to be stationary.
Because of the fluctuation-dissipation theorem, which assures the
system reaches an equilibrium asymptotically in time (steady state),
we have
\be
 {\bf C}(t - \tau) = k_B T {\bf K} (t - \tau) ,
 \label{autocorr}
\ee
where $k_B$ is the Boltzmann constant and $T$ is the surface
temperature.
Whenever this description is valid, Eq.~(\ref{autocorr}) is independent
of the particular details of the interaction between the system and the
surrounding heat bath.
For Ohmic friction (no memory effects), the kernel can be expressed as
a delta function and Eq.~(\ref{GLE}) becomes a standard Langevin
equation.

Former experimental and theoretical studies of Na diffusion on a
Cu(001) surface at low coverages were carried out by assuming that
the corresponding dynamics is well described by the motion of a single
adsorbate, governed by the standard Langevin
equation.\cite{ellis1,graham1}
The thermal effects induced by the surface temperature were accounted
for by a Gaussian white noise, this being in agreement with
contemporary full molecular dynamics simulations,\cite{ellis2} where
the motion of both the surface atoms and a sample of adatoms scattered
over the surface was taken into account.
As coverage increases, the Langevin treatment has still been used
in the literature,\cite{allison} but in combination with molecular
dynamics (Langevin Molecular Dynamics, LMD), where the inter-adsorbate
forces are commonly described by repulsive dipole-dipole
potentials.\cite{ying1,coverage1}
An interesting result obtained from this type of simulations is the
appearance of ordered structures at $\theta = 0.125$ and $\theta =
0.16$, compatible with a dominant repulsive interaction, as observed
experimentally.\cite{graham2}
Therefore, this fact should be taken into account as a limitation of
the range of validity for fully based stochastic approaches, since they
can not describe the appearance of this kind of collective phenomena
or phases.
Within the context of the GLE, the memory effects on the frictional
damping in surface diffusion have also been discussed\cite{ying3} by
considering a Hamiltonian similar to the Caldeira-Leggett
Hamiltonian.\cite{maga,caldeira,cortes}
This leads to an analysis in terms of the relationship between the
surface excitation frequency (i.e., the Debye frequency), $\omega_D$,
and the characteristic vibrational frequency of the adatoms, $\omega_0$.
In general, when $\omega_D$ is greater than $\omega_0$, memory effects
are not very important and, therefore, an Ohmic friction can be
assumed.
This is the case, for example, for Na atoms diffusing on a Cu(001)
surface, where the Debye frequency is about 30~meV, while the lowest
T-mode frequency is about 6~meV.

Though the GLE is a phenomenological equation, it can be formally
derived from a microscopic Hamiltonian\cite{maga,caldeira,cortes}
consisting of the system coupled to a bath of harmonic oscillators
(the one-bath model).
This formalism has been used, for example, to study quantum tunneling
in dissipative systems,\cite{caldeira,grabert,eli1} the so-called
Kramers' turnover problem,\cite{eli2,melnikov,eli3,eli4} and vibrational
dephasing rates.\cite{eli5}
A remarkable aspects of this formalism is that it allows us to
establish straightforwardly a connection between the Langevin formalism
in surface diffusion and quantum mechanical approaches, such as the one
introduced by Manson and Celli.\cite{manson}
Hence, since Eq.~(\ref{GLE}) is very general, we can also assume that
there are two independent, uncorrelated baths (one for the phonons and
another for the adsorbates) in order to describe surface diffusion
processes with interacting adsorbates within the framework of a
two-bath model based on a generalization of the one-bath
Caldeira-Legget Hamiltonian.
In this way, two independent fluctuating forces are present and the
total kernel then consists of the sum of the kernels associated with
each fluctuating force.
As mentioned, if each bath is assumed as Ohmic, both kernels will then
be describable in terms of delta functions in time.
This leads immediately to the so-called {\it interacting single
adsorbate} (ISA) {\it model},\cite{ruth1,ruth2,ruth3} which has been
shown to reproduce fairly well experimental surface diffusion data
under the presence of interacting adsorbates up to moderate
coverages\cite{cp} ($\theta \sim 0.12$).
Kramers' theory can also be generalized using two baths to infer
physical properties of the diffusing particle through the quasielastic
(Q) peak obtained from experimental data, as previously proposed within
the one-bath model.\cite{raul1,raul2}
Furthermore, the vibrational dephasing theory proposed for the
low-frequency frustrated translational motion or T-mode within the
one-bath model can also be straightforwardly extended.\cite{raul3}
This allows us to provide some analytical expressions, firmly based on
formal grounds, for a fitting procedure in order to extract
relevant information from the experimental data.

The organization of this work is as follows.
In Sec.~\ref{sec2} we present the basic building blocks of the
theoretical formalism which allow to connect the two-bath and the
ISA models, as well as their connection to the experiment.
In order to illustrate the applicability of this formalism, in
Sec.~\ref{sec3} we present an analysis of Na diffusion on Cu(001)
where a lot of experimental data are available from the
literature.\cite{coverage1,coverage2}
Finally, in Sec.~\ref{sec4} we summarize the main conclusions derived
from this work.

%%%%%%%%%%%%%%%%%%%%%%%%%%%%%%%%%%%%%%%%%%%%%%%%%%%%%%%%%%%%%%%%%%%%%%%
%%%%%%%%%%%%%%%%%%%%%%%%%%%%%%%%%%%%%%%%%%%%%%%%%%%%%%%%%%%%%%%%%%%%%%%

\section{Theory}
\label{sec2}

%%%%%%%%%%%%%%%%%%%%%%%%%%%%%%%%%%%%%%%%%%%%%%%%%%%%%%%%%%%%%%%%%%%%%%%

\subsection{Basic formalism}

In diffusion experiments carried out by means of QHAS, one measures
the differential reflection coefficient which, in analogy to
liquids,\cite{vanHove} reads as
\ba
 \frac{d^2 {\mathcal R} (\Delta {\bf K}, \omega)}{d\Omega d\omega}
 & = & n_d F S(\Delta {\bf K}, \omega)
  \nonumber \\
 & = & n_d F \iint G({\bf R},t)
  e^{i(\Delta {\bf K} \cdot {\bf R} -\omega t)}
   \ \! d{\bf R} \ \! dt . \nonumber \\ & &
 \label{eq:DRP}
\ea
This expression gives the probability that the probe He atoms scattered
from the diffusing collective (distributed upon the surface with a
concentration $n_d$) reach a certain solid angle $\Omega$ with an
energy exchange $\hbar\omega = E_f - E_i$ and wave vector transfer
parallel to the surface $\Delta {\bf K} = {\bf K}_f - {\bf K}_i$.
In Eq.~(\ref{eq:DRP}), $F$ is the atomic form factor, which depends on
the interaction potential between the probe atoms in the beam and the
adparticles on the surface.
On the other hand, $S(\Delta {\bf K},\omega)$ is the so-called
{\it dynamic structure factor} or {\it scattering law}.
Within the context of the linear response theory, this observable
corresponds to a {\it response function} and provides a complete
information about the dynamics and structure of the adsorbates
through particle distribution functions.
Experimental information about long-distance and long-time correlations
are obtained for small values of $\Delta K$ and small energy transfers
$\hbar \omega$, respectively, through the Q and T-mode peaks.
Particle distribution functions can be well-described by means
of the so-called van Hove or time-dependent pair correlation
function,\cite{vanHove} $G({\bf R},t)$, when dealing with
interacting particles.
Given a particle at the origin at some arbitrary initial time $t = 0$,
$G({\bf R},t)$ gives the averaged probability of finding the same or
another particle at the surface position ${\bf R}$ at a time $t$.
This function thus generalizes the well-known pair distribution
function $g({\bf R})$, commonly used within the context of statistical
mechanics,\cite{mcquarrie,hansen} by providing information about the
interacting particle dynamics.

Alternatively, the dynamic structure factor can be
expressed\cite{vanHove} as
\ba
 S(\Delta {\bf K},\omega)& = &
  \int e^{-i\omega t} \ \!
   \langle e^{-i\Delta {\bf K} \cdot {\bf R}(t)}
    e^{i\Delta {\bf K} \cdot {\bf R}(0)}\rangle \ \! dt
  \nonumber \\
   & = & \int e^{-i\omega t} \ \! I(\Delta{\bf K},t) \ \! dt ,
 \label{eq:DSF}
\ea
where the brackets in the integral denote an ensemble average.
In the second line of Eq.~(\ref{eq:DSF}),
\be
 I(\Delta {\bf K},t) \equiv
  \langle e^{-i\Delta {\bf K} \cdot
   [{\bf R}(t) - {\bf R}(0)] } \rangle
  = \langle e^{-i\Delta K \cdot
    \int_0^t v_{\Delta {\bf K}} (t') \ \! dt'} \rangle
 \label{eq:IntSF}
\ee
is the {\it intermediate scattering function} (also known as
{\it polarization}\cite{allison,farago,fouquet} in spin-echo
experiments), which is the space Fourier transform of $G({\bf R},t)$.
In Eq.~(\ref{eq:IntSF}), $v_{\Delta {\bf K}}$ is the velocity of
the adparticle projected onto the direction of the parallel momentum
transfer, $\Delta {\bf K}$.
The averages involved in Eq.~(\ref{eq:IntSF}) can be obtained from
Langevin numerical simulations for an adparticle in the presence of
an external field of force $V({\bf R})$, which gives the
substrate-adsorbate interaction, and two uncorrelated noise functions
leading to two friction coefficients: one associated with thermal noise
(surface phonons) and another with the collisional friction due to the
collisions among adsorbates.

%%%%%%%%%%%%%%%%%%%%%%%%%%%%%%%%%%%%%%%%%%%%%%%%%%%%%%%%%%%%%%%%%%%%%%%

\subsection{Kramers' theory for interacting adsorbates}

As mentioned above, Manson and Celli\cite{manson} proposed a quantum
diffuse inelastic theory for small and intermediate coverages
based on the transition matrix formalism, ignoring multiple scattering
effects with He atoms.
Within this approach, $S(\Delta {\bf K},\omega)$ is obtained after
assuming that all crystal vibrational modes ($N$) and point-like
scattering centers ($M$) satisfy the harmonic approximation with a
given frequency distribution. In a similar way, we can describe the
diffusion of interacting adsorbates by means of two types of
independent, uncorrelated baths.

Based on Kramers' theory,\cite{kramers,peter} a quantum and classical
theory of surface diffusion at very low coverages has been developed
in recent years.\cite{eli6,raul1,raul2}
Within this theory, the corresponding GLE is split up into two coupled
equations, one for each degree of freedom.
This GLE arises from a total system+bath Hamiltonian expressed in terms
of a single bath consisting of a set of $N$ harmonic oscillators.
This bath simulates the surface thermal effects on the adsorbate.
For low and moderate coverage ($\theta \lesssim 0.12$), a similar
model Hamiltonian but with two baths can also be formulated.
In this case, apart from the surface thermal effects, we take one
adsorbate as the tagged particle or system, while the remaining ones
constitute the second bath, which is described as a collection of $M$
harmonic oscillators.
Obviously, the characteristics of the second bath are going to be
dependent on the surface coverage considered in the experiment.
Thus, the corresponding total (system+two-bath) Hamiltonian reads as
\begin{eqnarray}
 H & = & \frac{p_x^2 +p_y^2 }{2m} + V(x,y) \nonumber \\
  & & + \frac{1}{2} \sum_{i=1}^{N}
 \left[ \frac{p_{x_i}^2}{m_i} + m_i
  \left( \omega_{x_i} x_i - \frac{c_{x_i}}
    {m_i \omega_{x_i}} \ \! x \right)^2  \right]
   \nonumber \\
  & & + \frac{1}{2} \sum_{i=1}^{N}
 \left[ \frac{p_{y_i}^2}{m_i}+ m_i
  \left( \omega_{y_i} y_i - \frac{c_{y_i}}
  {m_i \omega_{y_i}} \ \! y \right)^2 \right]
  \nonumber \\
     & & + \frac{1}{2} \sum_{j=1}^{M}
 \left[ \frac{{\bar p}_{x_j}^2}{{\bar m}_j} + {\bar m}_j
  \left( {\bar \omega}_{x_j} {\bar x}_j
  - \frac{d_{x_j}}{{\bar m}_j {\bar \omega}_{x_j}} \ \! x \right)^2
   \right]
  \nonumber \\
  & & + \frac{1}{2} \sum_{j=1}^{M}
 \left[ \frac{{\bar p}_{y_j}^2}{{\bar m}_j} + {\bar m}_j
  \left( {\bar \omega}_{y_j} {\bar y}_j
   - \frac{d_{y_j}}{{\bar m}_j {\bar \omega}_{y_j}} \ \! y \right)^2
   \right] ,
\label{HCL2}
\end{eqnarray}
where $V(x,y)$ is the periodic potential describing the substrate and
$(p_x,p_y)$ and $(x,y)$ are the adparticle momenta and positions
on the surface, and $(p_{x_i},x_i)$ and $(p_{y_i},y_i)$ with $i=1,
\cdots , N$ are the momenta and positions of the $i$th bath oscillator
(phonon), with masses and frequencies given by $m_i$ and $\omega_{k_i}$
($k = x, y$), respectively.
Phonons with polarization in the $z$-direction are not considered.
The same holds for the barred magnitudes, though they are associated
with an assumed harmonic oscillator bath of $M$ adsorbates, thus
neglecting eventual long-distance correlations.
The $c_{k_i}$ and $d_{k_i}$ coefficients in both directions ($k=x,y$)
give the adsorbate-substrate and adsorbate-adsorbate coupling
strengths, respectively.

The spectral density for the two baths is defined similarly to the case
of one bath,
\begin{eqnarray}
  J_k(\omega) & = & \frac{\pi}{2}
 \sum_{i=1}^N \frac{c_{k_i}^2}{m_i \omega_{k_i}^2}
  \ \delta (\omega - \omega_{k_i}) \nonumber \\
  & & + \frac{\pi}{2}
 \sum_{j=1}^M \frac{d_{k_j}^2}{{\bar m}_j {\bar \omega}_{k_j}^2}
  \ \delta (\omega - {\bar \omega}_{k_j}) , \qquad k=x,y,
 \label{SD2}
\end{eqnarray}
though now it contains two terms: one spectral density is associated
with the surface phonons and the other one with the bath of adsorbates.
Equation~(\ref{SD2}) enables passing to a continuum model provided
the time-dependent friction can be expressed as

\begin{equation}
 \eta_k (t) = \frac{2}{\pi}
  \int_0^\infty \cos \omega t \ \frac{J_k(\omega)}{\omega} \ d \omega ,
  \qquad k = x, y .
\end{equation}
In the case of Ohmic friction, i.e., $\eta (t) = 2 \eta \delta (t)$,
Eq.~(\ref{GLE}) reduces to the standard Langevin equation\cite{ruth1,%
ruth2,ruth3} (the delta function counts only one half when the
integration is carried out from $0$ to $\infty$),
\begin{equation}
 \ddot{\bf R} = - \eta \dot{\bf R} + {\bf F}({\bf R})
  + \delta {\bf F}_f ,
 \label{eq6}
\end{equation}
which constitutes the basis of the ISA model.
In this equation, $\delta {\bf F}_f$ is given by the sum of two
noncorrelated noises: the lattice (thermal) vibrational effects and the
adsorbate-adsorbate collisions, which are simulated by a Gaussian white
noise ($G$) and a white shot noise ($S$), respectively.
Thus, for each degree of freedom, we have $\delta F_f (t) =
\delta F_f^G (t) + \delta F_f^S (t)$.
The Gaussian white noise satisfies the properties: $\langle F_f^G (t)
\rangle = 0$ and $\langle F_f^G(t) F_f^G(t') \rangle = 2 m \gamma
k_B T \delta(t'-t)$, where $m$ is the adsorbate mass and $\gamma$ is
the (constant) friction coefficient measuring the adsorbate-phonon
coupling strength.
On the other hand, the shot noise is given\cite{ruth4} by a sum of
pulses mimicking the collision impacts.
In the Markovian approximation, these collisions are assumed to be
sudden (strong but elastic), with post-collision effects relaxing
exponentially at a constant rate much larger than the average number
of collision per time unit or {\it collisional friction}, $\lambda$.
The memory function or kernel associated with the shot noise in
(\ref{GLE}) then becomes local in time and $\delta F_f^S(t)$ will
display features of a {\it white} shot noise.\cite{ruth2}
In other words, the adsorbate-adsorbate collisions are
described by means of a white shot noise as a limiting case of a
colored shot noise. Thus, $K(t-t') \simeq 2 \eta \ \! \delta(t-t')$,
where the total (Ohmic) friction is $\eta = \gamma +
\lambda$. A simple relationship between the collisional friction,
$\lambda$, and the coverage, $\theta$, at a temperature $T$ is
given\cite{ruth4} by
\be
 \lambda = \frac{6 \rho \theta}{a^2} \sqrt{ \frac{k_B T}{m} } .
 \label{theta}
\ee

The theory of activated surface diffusion in one dimension was
developed from Kramers' solution\cite{eli4,eli6} to the problem
of escape from a metastable well.\cite{kramers,peter}
The underlying dynamics is assumed well described by the Langevin
equation provided that the reduced barrier height is of the order
of $\sim$3 or higher (i.e., $V^\ddagger / k_B T \gg 3$).
The energy loss to the bath of trajectories close to the barrier top is
given by classical mechanics and the potential at the barrier top is
approximately parabolic.
Kramers' based theory with finite barrier correction terms can
then be replaced by Langevin numerical simulations.
As is well known, the so-called {\it turnover region}\cite{eli3,peter}
is observed when the transmission factor or prefactor of the
exponential law for the escape rate is plotted versus the friction
coefficient at a given surface temperature.
Such a factor behaves linearly with $\eta$ for relatively low
frictions (i.e., $\sim \eta$) and goes like $\sim \eta^{-1}$ at
the low-to-moderate friction regime (Smoluchowski limit).
In between, it passes through a maximum, thus undergoing the turnover,
where it displays a characteristic smooth shape.

The starting point of the Kramers' model is a kinetic equation in one
dimension for the stationary flux of particles exiting each well at
either barrier.
This flux is affected by the rate of particles exiting the $j$th well
and those arriving at the well from the two neighboring wells, $j+1$
and $j-1$.
Here we are going to give the main analytical expressions derived
from Kramers' theory, but more details can be found
elsewhere.\cite{eli4,eli6,raul2}
A central quantity in the theory is the reduced average energy loss,
$\delta$, to the bath as the adatom traverses from one barrier to the
next. If a single cosine potential with a barrier height $V^\ddagger
= 2V_0$ is considered, the energy loss is given by
\be
 \delta = \frac{8 V_0 \eta}{k_B T \omega_0} ,
 \label{delta}
\ee
where $\omega_0 = 2 \pi \sqrt{V_0 /m a^2}$ is the harmonic frequency
of the oscillations near the well bottom, $m$ is the mass of the
adatom and $a$ is the unit cell length.
Since many experiments or calculations are carried out typically under
conditions of large reduced barrier heights, $\delta$ can be unity or
even larger, even though the damping constant is rather small.

In the moderate to strong friction regime, where the rate limiting
step is the spatial diffusion ($sd$) across the barrier, the rate
of the escape from the well in both directions is given by the
Kramers-Grote-Hynes formula,\cite{kramers,grote}
\be
 \Gamma_{sd} = \frac{\lambda^\ddagger}{\omega^\ddagger}
 \frac{\omega_0}{\pi} \ e^{-V^\ddagger /k_B T} ,
 \label{sd}
\ee
where the corresponding prefactor is
\be
 \frac{\lambda^\ddagger}{\omega^\ddagger} =
  \sqrt{1 + \left( \frac{\eta}{2\omega^\ddagger} \right)^2 }
   - \frac{\eta}{2\omega^\ddagger} ,
 \label{pref}
\ee
the magnitudes associated with the barrier being denoted by $\ddagger$.
This prefactor appears as a renormalization taking into account
recrossings, since we are working implicitly in normal mode
coordinates for the diffusing particle and the baths.\cite{eli3}
Finally, for the partial rates one finds
\begin{align}
 \Gamma_j & = - \frac{\Gamma_{sd}}{\pi} \int_{0}^{2\pi}
   \sin^2 \left( \frac{\Delta K}{2} \right) \cos(j \Delta K) \
   \nonumber \\
  & \times \exp \left\{ \frac{2}{\pi} \int_{0}^{\pi/2}
    \ln \left[ \frac{1 - P^2(x)}{1 + P^2(x) - 2 P(x) \cos(\Delta K)}
  \right]
   dx \right\} \nonumber \\
  & \times d\Delta K ,
 \label{gammaj}
\end{align}
where $P(x) = e^{- \delta/4\cos^2(x)}$.
The escape rate from the zeroth surface well is $\kappa = - \Gamma_0$
and the relative probability for a jump of length $j$ is given by the
probability of trapping at the $j$th well, $P_j = \Gamma_j/\kappa$.
For a one-dimensional periodic potential, the diffusion coefficient is
related to the escape rate as
\be
 D = \frac{1}{2} \ \kappa \langle l^2 \rangle
   = \frac{1}{2} \ a^2 \sum_{j=-\infty}^\infty j^2 \Gamma_j ,
\ee
where $\langle l^2 \rangle$ is the mean square path length.
By using Eq.~(\ref{gammaj}), this diffusion coefficient can be
reexpressed in a more compact form as
\be
 D = D_{sd} \Upsilon^{-1} \
  \exp \left\{\frac{2}{\pi} \int_{0}^{\pi/2}
   \ln \left[ 1 + P(x) \right] dx \right\} ,
\label{difc}
\ee
where $D_{sd} = (1/2) a^2 \Gamma_{sd}$ is the diffusion coefficient in
the spatial diffusion regime and $\Upsilon$ is the depopulation factor
for the metastable well,
\be
 \Upsilon = \exp \left\{ \frac{2}{\pi} \int_{0}^{\pi/2}
   \ln \left[ 1 - P(x) \right] dx \right\} ,
\ee
first given by Melnikov.\cite{melnikov}

Now, in analogy to other models, such as the Chudley-Elliott
model,\cite{chudley,ruth2} an analytical expression for the full width
at half maximum (FWHM) of the dynamic structure factor can also be
obtained by imposing a master equation for the intermediate scattering
function.\cite{raul2}
If the dynamic structure factor has a Lorentzian shape, the FWHM is
given by
\begin{align}
 & \Gamma(\Delta K) = 4 \Gamma_{sd}
  \sin^2 \left( \frac{\Delta K}{2} \right) \nonumber \\
 & \times \exp \left\{ \frac{2}{\pi} \int_{0}^{\pi/2}
   \ln \left[ \frac{1 - P^2(x)}{1 + P^2(x) - 2P(x)\cos(\Delta K)}
     \right] dx \right\} .
% \nonumber \\ & &
 \label{1.0}
\end{align}
This equation is important in the sense that, assuming the validity of
Kramers' model and the master equation approach mentioned above, it
allows for a direct comparison with the experimental data and/or
Langevin numerical simulations and therefore an estimation of the
spatial diffusion rate $\Gamma_{sd}$ and the energy loss $\delta$.
From these parameters and their temperature dependence, one can further
infer the barrier height, the friction coefficient and the barrier
frequency.

%%%%%%%%%%%%%%%%%%%%%%%%%%%%%%%%%%%%%%%%%%%%%%%%%%%%%%%%%%%%%%%%%%%%%%%

\subsection{Vibrational dephasing and T-mode}

The vibrational dephasing of an adsorbate on a surface arises from the
random frequency fluctuation undergone by the presence of the surface
bath.
Dephasing mainly emerges from the anharmonicity of the potential
describing the adsorbate vibrations.
When the influence of the coverage becomes important, another source
of fluctuations arises, which is associated with the interaction among
adsorbates.
Within the ISA model,\cite{ruth1,ruth2,ruth3} for example, this effect
is accounted for by a white shot noise.
Dephasing times or rates can be obtained from the autocorrelation
function of the frequency fluctuation through the Green-Kubo
expression.\cite{eli5}

In Ref.~\onlinecite{raul3}, simple analytic expressions are derived
for the vibrational line shapes of the T-mode from a Hamiltonian
analogous to (\ref{HCL2}), though in the case of the one-bath model.
In particular, these expressions describe the location and FWHM of the
T-mode peak as a function of the surface temperature.
According to such a model, to leading order, the T-mode potential of
mean force can be expressed generically as
\begin{equation}
 V(q) = \frac{1}{2} \ \omega _0^2 q^2 + K_4 q^4 ,
 \label{1.1}
\end{equation}
where $q$ is the mass weighted T-mode vibrational coordinate,
$\omega _0$ is the harmonic frequency, and $K_4$ is the anhamonicity
constant.
To second order in the anharmonicity and provided the damping constant
$\eta$ is not too large $(\eta/\omega _0 \leq 2)$, for Ohmic friction
one finds that the peak location, $\langle \omega_T\rangle$, is given
by
\begin{equation}
% \langle \omega _T \rangle = \sqrt{\omega_0^2 - \frac{\eta^2}{4}}
%  + k_B T \frac{6 K_4}{\omega_0^2
%   \sqrt{\displaystyle \omega_0^2 - \frac{\eta^2}{4}}}
%  \left( 1 - k_B T \frac{9 K_4}{\omega_0^4} \right) .
 \langle \omega_T \rangle = \omega_{\rm eff}
  + k_B T \ \frac{6 K_4}{\omega_0^2 \omega_{\rm eff}}
   \left( 1 - k_B T \frac{9 K_4}{\omega_0^4} \right) ,
 \label{1.2}
\end{equation}
where $\omega_{\rm eff} = \sqrt{\omega_0^2 - (\eta/2)^2}$ in the case
of the two-bath model described by (\ref{HCL2}).
Similarly, the T-mode peak FWHM, $\sigma$, depends on temperature and
friction as
\begin{equation}
 \sigma =
  k_B T \ \frac{6 \left| K_4 \right| }{\omega _0^2 \omega_{\rm eff}} .
 \label{1.3}
\end{equation}

%%%%%%%%%%%%%%%%%%%%%%%%%%%%%%%%%%%%%%%%%%%%%%%%%%%%%%%%%%%%%%%%%%%%%%%
%%%%%%%%%%%%%%%%%%%%%%%%%%%%%%%%%%%%%%%%%%%%%%%%%%%%%%%%%%%%%%%%%%%%%%%

\section{Results}
\label{sec3}

\begin{figure}
 \includegraphics[width=7cm]{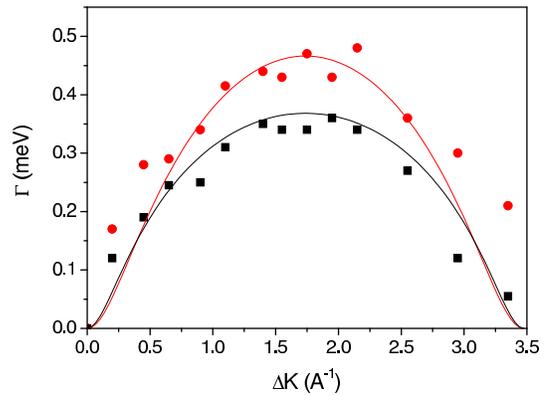}
 \caption{\label{fig1}
  FWHM as a function of the parallel momentum transfer for Na diffusion
  on Cu(001) along the $[100]$ direction at $T = 300$~K.
  Experimental data\cite{coverage1} are indicated with symbols, while
  the solid lines are the fitted curves, obtained by applying
  Eq.~(\ref{1.0}).
  Two different values of the coverage are considered: $\theta_1 =
  0.028$ (black) and $\theta_2 = 0.078$ (red).}
\end{figure}

Kramers' turnover theory provides a two-parameter description
of the diffusion process through the FWHM of the Q-peak versus
the parallel momentum transfer.
As seen in the preceding Section, these two parameters are the spatial
diffusion rate and the energy loss.
This treatment suits particularly well in fitting procedures applied
to experimental data and/or numerical Langevin-type results due to the
closed analytical expressions that are involved.
Very good estimates of rates, diffusion coefficients and jump
distributions are obtained when the barrier for diffusion exceeds
the thermal energy and finite barrier corrections are also included.
In Fig.~\ref{fig1}, experimental\cite{coverage1} (symbols) and
numerically fitted (solid lines) FWHM are plotted as a function of the
parallel momentum transfer $\Delta K$ for Na diffusion on Cu(001) along
the $[100]$ direction at $T = 300$~K.
Two different coverages are considered: $\theta_1 = 0.028$
(black squares/line) and $\theta_2 = 0.078$ (red circles/line).
Fittings have been carried out using Eq.~(\ref{1.0}).
The fitted values obtained for the spacial diffusion rate are
$\Gamma_{sd}^{(1)} = 0.250$~meV and $\Gamma_{sd}^{(2)} = 0.238$~meV,
and for the energy loss, $\delta_1 = 1.29$ and $\delta_2 = 1.97$.
The broadening of the Q-peak with coverage is quite well reproduced by
Eq.~(\ref{1.0}).
Even more, as is clearly seen from Eq.~(\ref{difc}), the diffusion
coefficient decreases with the coverage (as shown previously in
Ref.~\onlinecite{ruth3}) and the energy loss increases with the
coverage or friction according to Eq.~(\ref{delta}).

\begin{figure}
 \includegraphics[width=7cm]{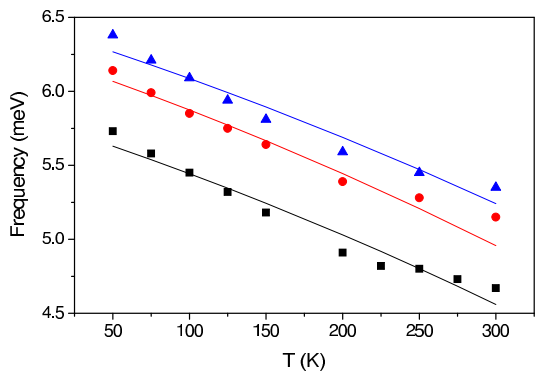}
 \caption{\label{fig2}
  T-mode frequency as a function of the surface temperature for Na
  diffusion on Cu(001) along the direction $[100]$.
  Experimental data\cite{coverage2} are indicated with symbols, while
  the solid lines are the fitted curves, obtained by applying
  Eq.~(\ref{1.2}).
  Three different values of the coverage are considered: $\theta_1 =
  0.028$ (black), $\theta_2 = 0.078$ (red) and $\theta_3 = 0.125$
  (blue).}
\end{figure}

\begin{figure}
 \includegraphics[width=7cm]{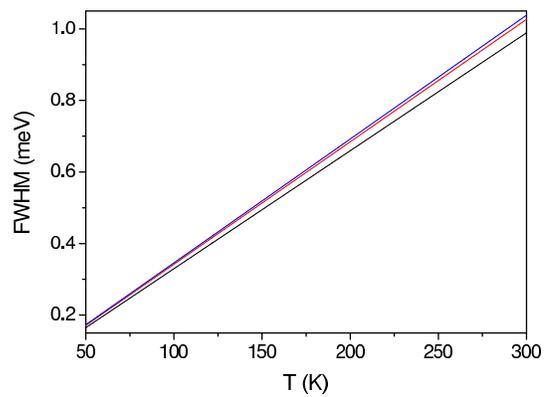}
 \caption{\label{fig3}
  T-mode FWHM as a function of the surface temperature for Na diffusion
  on Cu(001) along the direction $[100]$.
  The solid lines have been obtained by means of Eq.~(\ref{1.3})
  and using the values arisen from the fittings displayed in
  Fig.~\ref{fig2}.
  Three different values of the coverage are considered: $\theta_1 =
  0.028$ (black), $\theta_2 = 0.078$ (red) and $\theta_3 = 0.125$
  (blue).}
\end{figure}

The T-mode peak position and its FWHM are plotted in Figs.~\ref{fig2}
and \ref{fig3}, respectively, as a function of the surface temperature
and for three coverages: $\theta_1 = 0.028$ (black), $\theta_2 = 0.078$
(red) and $\theta_3 = 0.125$ (blue).
In Fig.~\ref{fig2}, the best fitted curves (solid lines) with respect
to the corresponding experimental data\cite{coverage2} (symbols) for
the T-mode peak positions have been obtained from Eq.~(\ref{1.2}).
The corresponding frequencies obtained when Eq.~(\ref{1.2}) is applied
are: $\omega_0^{(1)} = 2.13 \times 10^{-4}$, $\omega_0^{(2)} =
2.3 \times 10^{-4}$, and $\omega_0^{(3)} = 2.4 \times 10^{-4}$, all
given in atomic units (the nominal value being $2.20 \times 10^{-4}$
a.u.). With the coverage, the T-mode frequency shifts towards higher
values.
After the definition of the effective frequency in Eq.~(\ref{1.2}), it
should display the opposite trend with the coverage since a higher
friction coefficient is required in the ISA model.
However, the changes on $\omega_0$ are so small (less than 10$\%$)
that we can not rule out small changes in the mean force potential.
On the other hand, the best fitted values found for the anharmonicity
constants (also through Eq.~(\ref{1.2})) are $K_4^{(1)} = -6.17 \times
10^{-14}$, $K_4^{(2)} = -8.05 \times 10^{-14}$, and $K_4^{(3)} =
-9.25 \times 10^{-14}$, also in atomic units.
%The $\omega_0$ frequency ranges from $2.13 \times 10^{-4}$ to
%$2.4 \times 10^{-4}$ atomic units and the anharmonicity constant,
%$K_4$, varies from $- 6.17 \times 10^{-14}$ to $- 9.25 \times 10^{-14}$,
%for increasing coverage (see Table~\ref{tab1}).
For a cosine potential, $K_4$ is negative and therefore the
anharmonicity produces a red shift with a linear temperature
dependence.\cite{raul3}
The agreement with the surface temperature is fairly good when
second-order corrections in the anharmonicity are included.
The different $\omega_0$ and $K_4$ values are slightly modified due
to coverage, thus indicating a modification of the barrier/well and
anharmonicity due to increasing amount of adsorbates, respectively.
In general, in the presence of only one bath, the new system plus bath
frequencies are renormalized due to the surface friction or coupling
with the substrate.\cite{eli5}
This renormalization takes into account the sum over all harmonic
oscillators of the bath.
In the two bath scenario, the corresponding renormalization has an
extra sum coming from the adsorbates.
Finally, Fig.~\ref{fig3} remains as a prediction since no experimental
information is available about FWHM of the T-mode peaks.
However, what is well-known from experiments is that the T-mode peak
broadens with the coverage, which is theoretically predicted by
Eq.~(\ref{1.3}).

%\begin{table}
% \caption{Best fit frequencies and anharmonicity constants obtained
%  when Eq.~(\ref{1.2}) is applied to the experimental data displayed
%  in Fig.~\ref{fig2} for Na diffusion on Cu(001) along the direction
%  $[100]$ at three different coverages.
%  Both magnitudes are expressed in atomic units.}
% \begin{tabular}{c c c}
%  \hline\hline
%   $\theta$ & $\omega_0$ & $K_4$ \\
%   %heading
%   \hline
%   0.029 & \ 2.13$\times$10$^{-4}$ & -6.17$\times$10$^{-14}$ \\
%   0.078 & 2.3$\times$10$^{-4}$ & -8.05$\times$10$^{-14}$ \\
%   0.125 & 2.4$\times$10$^{-4}$ & -9.25$\times$10$^{-14}$ \\
%   \hline
% \end{tabular}
% \label{tab1}
%\end{table}

%%%%%%%%%%%%%%%%%%%%%%%%%%%%%%%%%%%%%%%%%%%%%%%%%%%%%%%%%%%%%%%%%%%%%%%
%%%%%%%%%%%%%%%%%%%%%%%%%%%%%%%%%%%%%%%%%%%%%%%%%%%%%%%%%%%%%%%%%%%%%%%

\section{Conclusions}
\label{sec4}

The main physical consequences of the approach presented here are that,
on average, the tagged adparticle is seen as a diffusing particle among
adsorbates which modify effective barriers and wells leading to the
renormalization of frequencies as showed from a normal mode
analysis.\cite{eli1,eli2}
The corresponding renormalized frequencies have been reported for the
one-bath model elsewhere.\cite{eli2}
A straightforward generalization of the effective frequencies can be
done for the two-bath model.
In particular, this new approach could provide a theoretical framework
to better interpret the coverage dependence of tunneling diffusion
experiments\cite{zhu} within a quantum Markovian formalism.\cite{ruth5}
Many times, in surface diffusion LMD simulations surface barriers
are only changed with the coverage in order to reproduce experimental
data, keeping the corresponding frequencies fixed.

%%%%%%%%%%%%%%%%%%%%%%%%%%%%%%%%%%%%%%%%%%%%%%%%%%%%%%%%%%%%%%%%%%%%%%%

\acknowledgments

This work has been supported by the Ministerio de Ciencia e
Innovaci\'on (Spain) under Project FIS2007-62006 and a Sabbatical
(G.~Rojas-Lorenzo) with Ref.~SB2006-0011.
R.~Mart\'{\i}nez-Casado acknowledges the Royal Society for a Newton
Fellowship.
A.S.~Sanz acknowledges the Consejo Superior de Investigaciones
Cient\'{\i}ficas for a JAE-Doc contract. We also highly appreciate
the very constructive and positive comments made by one of the
referees.

%%%%%%%%%%%%%%%%%%%%%%%%%%%%%%%%%%%%%%%%%%%%%%%%%%%%%%%%%%%%%%%%%%%%%%%
%%%%%%%%%%%%%%%%%%%%%%%%%%%%%%%%%%%%%%%%%%%%%%%%%%%%%%%%%%%%%%%%%%%%%%%

\end{document}